\documentstyle[12pt]{article}

\newcommand{\beq}{\begin{equation}}
\newcommand{\eeq}{\end{equation}}
\newcommand{\beqa}{\begin{eqnarray}}
\newcommand{\eeqa}{\end{eqnarray}}
\newcommand{\beqar}{\begin{eqnarray*}}
\newcommand{\eeqar}{\end{eqnarray*}}

\newcommand{\eg}{{\it e.g.,}\ }
\newcommand{\ie}{{\it i.e.,}\ }

\newcommand{\hphi}{\hat{\phi}}
\newcommand{\ha}{\hat{a}}
\newcommand{\hg}{\hat{\gamma}}
\newcommand{\bphi}{\bar{\phi}}
\newcommand{\ba}{\bar{a}}
\newcommand{\bg}{\bar{\gamma}}

\newcommand{\td}{\tilde{d}}
\newcommand{\tF}{\tilde{F}}
\newcommand{\tH}{\tilde{H}}
\newcommand{\tp}{\bar{p}}

\begin{document}
\begin{titlepage}
\rightline{\small hep-th/9512061 \hfill CERN-TH/95-331}
\rightline{\small \hfill McGill/95-61}
\vskip 5em

\begin{center}
{\bf \huge Rusty Scatter Branes}
\vskip 3em

{\large Ramzi R. Khuri$^{\rm a,b,}$\footnote{khuri@nxth04.cern.ch}}
and {\large Robert C.
Myers$^{\rm b,}$\footnote{rcm@hep.physics.mcgill.ca}}
\vskip 1em

{\em    $^{\rm a}\,$Theory Division, CERN, CH-1211\\
        Geneva 23, Switzerland}
\vskip 1.5em
{\em	$^{\rm b}\,$Department of Physics, McGill University \\
	Montr\'eal, Qu\'ebec, Canada H3A 2T8}
\vskip 4em

\begin{abstract}
We derive double dimensional reduction/oxidation in a framework
where it is applicable to describe general non-static
(and anisotropic) $p$-brane solutions.
Given this procedure, we are able
to relate the dynamical interaction potential for parallel extremal
$p$-branes in $D$ dimensions to that for extremal black holes in
$D-p$ dimensions. In particular, we find that to leading order
the potential vanishes for all $\kappa$-symmetric $p$-branes.
\end{abstract}
\end{center}

\vskip 3em

\rightline{\small CERN-TH/95-331\hfill}
\rightline{\small McGill/95-61\hfill}
\rightline{\small December 1995\hfill}

\end{titlepage}

\section{Introduction}

Recent investigations in nonperturbative string theory are beginning to
reveal new connections between different string theories \cite{conn}.
In fact many different string theories
appear to be different phases of a single underlying theory
\cite{mtheory}. Within these discussions, it appears that extended
objects, other than just strings, play an important role in
establishing the full web of connections relating various string
theories. Hence these exciting new results have generated a renewed
interest in $p$-branes (\ie $p$-dimensional extended objects)
and their interactions.

Many different background field solutions have been constructed
describing extended objects of different dimensions in various
theories (see \cite{report} and references
therein). Typically these solutions involve a single
$d$-form potential and a scalar field, \ie dilaton, coupled
to gravity. Further,
the solutions may be characterized as being static
and isotropic \cite{andy,ddduff} --- in fact, the latter is used to
refer to the fact that the solutions are Lorentz invariant in the
directions parallel to the world-volume of the $p$-brane.
With a $d$-form potential in $D$ dimensions,
one naturally finds two dual isotropic
$p$-branes. The first with $p=d-1$ carries an ``electric'' charge
from the $(d+1)$-form field strength. The dual
object with $p=D-d-3$ carries an analogous ``magnetic'' charge.

In certain cases, the $p$-branes
satisfy an extremality or zero-force condition, and
then two (or more) parallel $p$-branes can sit in static equilibrium.
In the latter case, though, one expects
that the precise cancellation of forces no longer holds when the
$p$-branes are in motion, and so they will scatter in a nontrivial
way. Only in very special cases would the scattering be trivial.
In the past,
there have been limited investigations of such scattering for
particular $p$-branes \cite{monscat,stscat,fbscat}.
The present paper provides a general analysis for the scattering of
$p$-branes in any dimension for a broad class of theories.

Our approach begins by
generalizing the usual reduction/oxidation procedure which
relates $p$-branes
in $D$ dimensions to $(p+1)$-branes in $D+1$ dimensions,
as described in section \ref{isox}. Our
discussion is quite general and in particular not restricted to
static or isotropic solutions. Thus this procedure can
accomodate a discussion of moving (parallel) $p$-branes, as required
to investigate scattering. As briefly reviewed in section
\ref{scat}, the scattering of extremal $p$-branes is determined by
a computation of the metric on the moduli space. However, combining
these calculations with our oxidation procedure, we can relate
scattering of $p$-branes to that of extremal black holes. Thus
the metric on the moduli space of extremal $p$-branes is simply
related to that calculated by Shiraishi \cite{shirscat} for extremal
black holes.
We conclude with a discussion of our results in section \ref{discuss}.

\section{Reduction and Oxidation} \label{isox}

Double-dimensional reduction is a procedure relating one class of
$(p+1)$-branes in a $(D+1)$-dimensional theory to another of $p$-branes
in $D$ dimensions. The inverse procedure going from $D$ to $D+1$
dimensions has come to be known as ``oxidation''.
Double-dimensional reduction was originally used to relate the
$D=11$ supermembrane action to that for the Type IIa superstring
in $D=10$ \cite{eleven}. The procedure was extended to generate new
solutions in general dimensions
in Refs.~\cite{ddduff,stain}, but their analysis was limited to
certain supersymmetric (static and isotropic)
solutions. Here our derivation of double
dimensional reduction provides a straightforward extension of the previous
discussions in that it is relevant for non-static $p$-brane solutions
in theories with an arbitrary dilaton coupling constant.
Such solutions will be relevant in the following discussion
of $p$-brane scattering.
We also point out that this procedure can be used to generate
anisotropic $p$-brane solutions, which do not have the usual
Lorentz invariance in the brane directions. As an example, the
most general extremal anisotropic brane solution of the
class of actions
described below is presented in Appendix \ref{app}.

The basic approach is to start with a $(p+1)$-brane solution
in $D+1$ dimensions, and eliminate one of the directions
parallel to the brane to produce a $p$-brane in
$D$ dimensions. We begin with the action:
\beq
I={1\over16\pi\hat{G}}\int d^{D+1}\!x\,\sqrt{-\hat{g}}\left[\hat{R}
-{\hg\over2}(\nabla\hat{\phi})^2
-{1\over 2(d+2)!}e^{-\ha\hg\hat{\phi}}\hat{F}^2
\right]
\label{iniact}
\eeq
where $\hat{\gamma}=2/(D-1)$.\footnote{We choose
a slightly unconventional normalization for the dilaton kinetic
term because it provides some simplification of the formulae
appearing in the discussion of $p$-brane scattering.}
Here $\hat{F}$ is a $(d+2)$-form field strength defined in terms
of a $(d+1)$-form potential $\hat{A}$
--- \ie $\hat{F}=d\hat{A}$. In the following, it will be useful to
define $\td=D-d-2$. Since we wish to allow
the consideration of anisotropic branes, we will not
restrict our discussion to the choices $p=d-1$ or $\td-2$
for the $(p+1)$-branes in $D+1$ dimensions.

We require that in the solutions all of the fields are independent
of (at least) one of the spatial coordinates, denoted $y$, which
runs parallel to the $(p+1)$-brane. This coordinate will be the
direction which is removed in the double-dimensional reduction.
Further we will require that in the $D+1$ dimensional solution
all of the
nonvanishing components of the $\hat{A}$ potential carry a $y$
index. In the dimensionally reduced
theory, we will have a $d$-form potential $A$ with
\[
A_{\mu\cdots\nu}=\hat{A}_{\mu\cdots\nu y}
\]
and a corresponding $(d+1)$-form field strength $F=dA$.
We also make a Kaluza-Klein reduction of the metric
which is required to have the form
\[
\hat{g}_{\mu\nu}=\pmatrix{\bar{g}_{\mu\nu}&0\cr
                          0&\exp[2\rho]\cr}
\]
where $\exp[2\rho]$ is the component $\hat{g}_{yy}$.
One finds that $\hat{R}(\hat{g})=\bar{R}(\bar{g})-2{\bar{\nabla}}^2\rho
-2(\bar{\nabla}\rho)^2$. Thus the gravity part of the action becomes
\[
\int d^{D+1}\!x\, \sqrt{-\hat{g}}\,\hat{R}= L
\int d^{D}\!x\, \sqrt{-\bar{g}}\,e^\rho\bar{R}
\]
where $L$ is the volume of the compactified dimension.
To remove the exponential factor in this action, we make a further
conformal transformation:
$\bar{g}_{\mu\nu}=\exp[-2\rho/(D-2)]g_{\mu\nu}$.
The Ricci scalar then becomes
\[
\bar{R}=\exp\left[{2\rho\over D-2}\right]\left(R
+{D-1\over D-2}\nabla^2\rho-{D-1\over D-2}(\nabla\rho)^2\right)
\]
and now the full dimensionally reduced action is
\[
I={L\over16\pi\hat{G}}\int d^D\!x\,\sqrt{-g}\left[R-{D-1\over
D-2}(\nabla\rho)^2-{\hg\over2}(\nabla\hat{\phi})^2
-{1\over2(d+1)!}e^{-\ha\hg\hat{\phi}}e^{-2\td\rho/(D-2)}F^2\right]\ .
\]

{}From this action, the scalar field equations of motion are
\beqar
2{D-1\over D-2}\nabla^2\rho+{2\td\over D-2}
\left({1\over2(d+1)!}e^{-\ha\hg\hat{\phi}}e^{-2\td\rho/(D-2)}F^2
\right)&=&0\\
\nabla^2\hat{\phi}+\ha\left(
{1\over2(d+1)!}e^{-\ha\hg\hat{\phi}}e^{-2\td\rho/(D-2)}F^2
\right)&=&0\ .
\eeqar
Combining these equations, we find that
$\nabla^2(\hat{\phi}-(D-1)\ha/\td\ \rho)=0$, and so this linear
combination represents a free scalar field which we have the liberty
to set to zero. Hence setting
$\rho=[\td/((D-1)\ha)]\hat{\phi}$ in the above action, we find
the kinetic term of the remaining scalar has an unconventional
normalization. Thus we scale this field to define the canonical
dilaton $\phi$ of the dimensionally reduced theory
\[
{\gamma\over2}(\nabla\phi)^2=
\left({\hg\over2}+{\td^2\over(D-2)(D-1)\ha^2}\right)(\nabla\hat{\phi})^2
\]
where $\gamma=2/(D-2)$.
With these choices the final action becomes
\beq
I={1\over16\pi G}\int
d^D\!x\,\sqrt{-g}\left[R-{\gamma\over2}(\nabla\phi)^2-{1\over2(d+1)!}
e^{-a\gamma\phi}F^2\right]
\label{finact}
\eeq
where $a^2=[(D-2)/(D-1)]\ha^2+\td^2/(D-1)$
and $G=\hat{G}/L$.

We can reverse this reduction process to construct an
oxidation prescription as follows: In the $D$-dimensional
theory described by the final action (\ref{finact}), we begin
with a $p$-brane solution with a given field configuration
\beq
g_{\mu\nu},\ \ \phi,\ \ A\ .
\label{oxaone}
\eeq
We add an extra dimension $y$ to
construct a $(p+1)$-brane solution for the initial
$(D+1)$-dimensional action (\ref{iniact}) as
\beq
\hat{A}_{\mu\cdots\nu y}=A_{\mu\cdots\nu}, \qquad\qquad
{\hphi/\ha}={\phi/a}
\label{oxatwo}
\eeq
\beq
\hat{g}_{\mu\nu}=\pmatrix{\exp\left[-{2\td\over(D-2)(D-1)}{\phi
\over a}\right]
g_{\mu\nu}&0\cr
0&\exp\left[{2\td\over(D-1)}{\phi\over a}\right]\cr}
\label{oxathree}
\eeq
with $\ha^2=[(D-1)/(D-2))]a^2 - \td^2/(D-2)$.

Applying Poincar\'e duality to the field strengths in the
above construction, one arrives at a distinct reduction/oxidation
scheme. First one replaces the $\hat{F}$ in the action
(\ref{iniact}) by the dual $(\td+1)$-form
field strength $\hat{H}=e^{-\ha\hg\hat{\phi}}\hat{\tF}$,
which then satisfies $d\hat{H}=0=d(e^{\ha\hg\hat{\phi}}\hat{\tH})$.
Hence the new field strength can be defined in terms
a $\td$-form potential $\hat{B}$, \ie $\hat{H}=d\hat{B}$,
and the new equations of motion will arise from the following
action
\[
I={1\over16\pi\hat{G}}\int d^{D+1}\!x\,\sqrt{-\hat{g}}\left[\hat{R}
-{\hg\over2}(\nabla\hat{\phi})^2
-{1\over 2(\td+1)!} e^{\ha\hg\hat{\phi}}\hat{H}^2
\right]\ .
\]
In the dimensional reduction with the dual field,
any components of $\hat{B}$ carrying a $y$ index will vanish,
\ie, $\hat{B}_{\mu\cdots\nu y}=0$. Further
in the reduced theory, we have a $\td$-form potential $B$ given by
\[
B_{\mu\cdots\nu}=\hat{B}_{\mu\cdots\nu}\ ,
\]
and a corresponding $(\td+1)$-form field strength $H=dB$.
The remainder of the construction is unchanged.
The essential point though is that one arrives at a second
independent reduction/oxidation procedure in which the form
potential is completely unchanged.

Expressed in terms of the action (\ref{finact}), the second
oxidation prescription is as follows: We begin in the $D$-dimensional
theory described by the final action (\ref{finact}) with a
$p$-brane solution given by the field configuration
\beq
g_{\mu\nu},\ \ \phi,\ \ A\ .
\label{oxbone}
\eeq
We add an extra dimension $y$ to
construct a $(p+1)$-brane solution for a
$(D+1)$-dimensional theory with the action
\beq
I={1\over16\pi\hat{G}}\int d^{D+1}\!x\,\sqrt{-\hat{g}}\left[\hat{R}
-{\hg\over2}(\nabla\hat{\phi})^2
-{1\over 2(d+1)!}e^{-\ha\hg\hat{\phi}}{F}^2
\right]
\label{finacttwo}
\eeq
where $\ha^2=[(D-1)/(D-2))]a^2 - d^2/(D-2)$.
Also ${F}=d{A}$ is the same $(d+1)$-form field strength that
appears in the original action. The field configuration of
the $(p+1)$-brane solution leaves the nonvanishing
components of the $d$-form potential $A$ unchanged, and has
\beq
{\hphi/\ha}={\phi/a}
\label{oxbtwo}
\eeq
\beq
\hat{g}_{\mu\nu}=\pmatrix{\exp\left[{2d\over(D-2)(D-1)}{\phi
\over a}\right]
g_{\mu\nu}&0\cr
0&\exp\left[-{2d\over(D-1)}{\phi\over a}\right]\cr}\ .
\label{oxbthree}
\eeq

Thus beginning with $p$-brane solutions of one theory (\ref{finact}),
we have two independent oxidation procedures.
The first prescription in eqs.~(\ref{oxaone}-\ref{oxathree})
extends the form degree of the potential $A$, as well as the dimension
of the brane
and the spacetime, while the second in eqs.~(\ref{oxbone}-\ref{oxbthree})
leaves the degree of the potential unchanged. Of course, the two
oxidation procedures lead to different $(D+1)$-dimensional
actions.

If one repeats either of the oxidation procedures twice adding two
extra dimensions, $y$ and $z$,
the final $(p+2)$-brane will isotropic in the two extra
dimensions,
\ie $\tilde{g}_{yy}=\tilde{g}_{zz}$ where $\tilde{g}_{\mu\nu}$
is the metric in $D+2$ dimensions.
Of course, this does not guarantee that the complete $(p+2)$-brane
is isotropic in all of its spatial directions. The latter would depend
on the details of the initial $p$-brane.
It is the case that if one begins with an isotropic $(d-1)$-brane
in $D$ dimensions with an electric ${F}$ charge and applies the
first oxidation procedure (\ref{oxaone}-\ref{oxathree}),
the resulting $d$-brane in $D+1$ dimensions and the
$(d+1)$-brane in $D+2$ dimensions are also isotropic \cite{report}.
Hence electrically charged isotropic $p$-branes for theories with arbitrary
dilaton coupling are
readily constructed by
beginning with a zero-brane or a point-like object in $D$ dimensions
carrying a conventional electric charge from a $U(1)$
two-form field strength,
and applying the oxidation procedure $p$ times to produce
a $p$-brane in $D+p$ dimensions --- see for example Appendix \ref{app}
and also
Refs.~\cite{ddduff,stain}. Similarly using the second
oxidation procedure (\ref{oxbone}-\ref{oxbthree}) and beginning
with black holes carrying a $(D-2)$-form magnetic charge
in $D$ dimensions, one can generate the conventional magnetically
charged isotropic $p$-branes.

If one applies the first oxidation scheme followed by the second
in adding two extra dimensions, $y$ and $z$,
the final $(p+2)$-brane is not isotropic in the two extra
dimensions,
\ie $\tilde{g}_{yy}\ne\tilde{g}_{zz}$.
Therefore multiple applications of both of these two prescriptions
result in not only isotropic but also anisotropic
membranes. As an example, Appendix \ref{app} provides the
most general extremal anisotropic brane solution for an action
of the form (\ref{finact}).

A further comment is that in our discussion neither oxidation
procedure makes any particular reference to the details of the
form of the original solution in eq.~(\ref{oxaone}) or (\ref{oxbone}).
In particular then, there is no requirement
that the initial solution be static. Thus
one could extend the preceding discussion so that if one begins with
a solution describing point-like objects in motion,
they would be oxidized
to $p$-brane solutions now moving in directions orthogonal to
the surfaces of the branes. This observation will be essential
in the discussion of $p$-brane scattering in the following
section.

\subsection{Mass and Charge Densities} \label{masse}

It is of interest to examine how the asymptotic physical properties
of the $p$-branes are affected in the oxidation procedures.
Since the oxidized solutions are independent of the extra
coordinate, their mass and charge densities will be simply
related to those of the original solutions. In both
prescriptions, the oxidation of the form potential is relatively
trivial, hence it is straightforward to show that the corresponding
charge densities are in fact unchanged. For example, applying the first
oxidation procedure (\ref{oxaone}-\ref{oxathree})
to a $p$-brane solution with electric charge density
$Q=\oint \tF$, results in a $(p+1)$-brane whose electric charge density
is given by the same integral, \ie $\hat{Q}=\oint \hat{\tF}
=\oint \tF=Q$. In other words, the charge density of the oxidized
solution is given by precisely the same parameters (or combination
of parameters) as that of the original solution.

A more interesting analysis is that of the mass density.
Suppose that we begin with a stationary $p$-brane solution for the action
(\ref{finact}) in $D$ dimensions. We assume that we have ``Cartesian''
coordinates on the background geometry
\[
x^\mu=(t,x^i,y^a)
\]
where $t$, $x^i$ and $y^a$ are time, $\tp$ transverse coordinates and
$p$ parallel coordinates for the brane, respectively ---
hence $D=p+\tp+1$. We assume that the
metric is independent of $t$ and $y^a$, which leads to a constant
mass density. The ADM mass per unit $p$-volume is
defined as follows \cite{massy}:
Asymptotically for $r^2=\sum_{i=1}^{\tp} (x^i)^2\rightarrow\infty$,
the metric is essentially
flat and so we define the deviation from a flat space as $h_{\mu\nu}=
g_{\mu\nu}-\eta_{\mu\nu}$. Now using the Cartesian coordinate
metric, the mass per unit $p$-volume $M_p$ is given by
\[
M_p
={1\over16\pi G}\oint \sum_{i=1}^{\tp}\ n^i\left[
\sum_{j=1}^{\tp}(\partial_jh_{ij}-\partial_i h_{jj})-
\sum_{a=1}^p\partial_i h_{aa}\right] r^{\tp-1} d\Omega
\]
where $n^i$ is a radial unit vector. If the solution is
oxidized following the first prescription, we find from
eq.~(\ref{oxbthree})
\[
\hat{h}_{\mu\nu}=\pmatrix{h_{\mu\nu}-{2\td\over(D-2)(D-1)}{\phi
\over a}\eta_{\mu\nu}&0\cr
0&{2\td\over(D-1)}{\phi\over a}\cr}
\]
where we assume that $\phi\rightarrow0$ asymptotically.
Hence the mass density of the resulting $(p+1)$-brane is
\beqar
M_{p+1}&=&{1\over16\pi\hat{G}}\oint \sum_{i=1}^{\tp}\ n^i
\left[\sum_{j=1}^{\tp}(\partial_j\hat{h}_{ij}
-\partial_i\hat{h}_{jj})-
\sum_{\ha=1}^{p+1}\partial_i\hat{h}_{\ha\ha}\right] r^{\tp+1}
d\Omega\\
&=&{1\over16\pi\hat{G}}\oint \sum_{i=1}^{\tp}\ n^i
\left[\sum_{j=1}^{\tp}\left(\partial_j\hat{h}_{ij}
-\partial_i\hat{h}_{jj}-{2\td\over(D-2)(D-1)}(
\delta_{ij}\partial_j\phi/a-\delta_{jj}\partial_i\phi/a)\right)
\right.\\
&&\hphantom{16\pi{G}\oint \sum_{i=1}^{\tp}\ n^i[\sum_{j=1}}
\left.-\sum_{a=1}^{p}\left(\partial_i\hat{h}_{aa}
-{2\td\over(D-2)(D-1)}\delta_{aa}\partial_i\phi/a\right)
-{2\td\over D-1}\partial_i\phi/a\right] r^{\tp+1}
d\Omega\\
&=&{1\over16\pi G\,L}\oint \sum_{i=1}^{\tp}\ n^i
\left[\sum_{j=1}^{\tp}(\partial_jh_{ij}-\partial_i h_{jj})-
\sum_{a=1}^{p}\partial_i h_{aa}\right] r^{\tp+1} d\Omega\\
&=& M_p/L\ \ .
\eeqar
All of the explicit $\phi$ contributions cancel out in $M_{p+1}$.
It is straightforward to show that the same cancellation arises when
applying the second oxidation procedure. Thus in both cases, the
parameters describing the mass density of the $p$-brane and
its oxidized counterpart
are identical irrespective of the details of the solutions.

One might also define $Q_D$, a
dilaton charge density for a $p$-brane. A simple definition for
this scalar charge density would be in terms of the
asymptotic behavior of the dilaton
\[
\phi\simeq -{Q_D\over r^{\tp-2}}
\]
again assuming that $\phi\rightarrow 0$. Since both oxidation procedures
yield $\hat{\phi}/\ha=\phi/a$, the above
definition yields $\widehat{Q}_D=(\ha/a)\,Q_D$.

\section{Scattering} \label{scat}

The existence of static multi-soliton solutions, including
multi-extreme black hole and $p$-brane solutions,
relies on the cancellation of the exchange forces generated by
the scalar, form-potential and gravitational fields
(the so-called ``zero-force''
condition). If the solitons are given velocities, however,
the zero-force condition ceases to hold and dynamical,
velocity-dependent forces arise. The full time-dependent equations
of motion that result are highly nonlinear and in general very
difficult to solve. In the absence of exact time-dependent
multi-soliton solutions, Manton's procedure \cite{manton}
for the computation of a metric on the soliton moduli space
yields a good low-velocity approximation for their exact dynamics.
Manton's method
may be summarized as follows: One begins with a static
multi-soliton solution, and gives time-dependence to
the moduli characterizing the configuration. One then calculates $O(v)$
corrections to the fields by solving the constraint
equations of the system with time-dependent moduli.
The resulting time-dependent field configuration only
satisfies the full field equations to lowest order in the
velocities, and so neglects the effects of any radiation fields.
Such a configuration would provide the
initial data of fields and time derivatives for an exact solution.
Another way of saying this is that the initial motion is
tangent to the set of exact static solutions. An effective
action valid to $O(v^2)$ describing the soliton motion is constructed
by substituting the solution to the constraints into the field theory
action. The resulting kinetic action
defines a metric on the moduli space of static solutions, and
the geodesic
motion on this metric approximates the dynamics of the solitons.
This approach was first applied to study the scattering of
BPS monopoles \cite{manton}, and a complete calculation of the
corresponding
moduli space metric and a description of its geodesics was
worked out by Atiyah and Hitchin \cite{atiyah}.
Manton's method was subsequently adapted to general
relativity by Ferrell and Eardley \cite{fere} for the study of
low-velocity scattering of extreme Reissner-Nordstrom black holes.

More recently, Shiraishi \cite{shirscat} adapted the method of
Ref.~\cite{fere} to obtain the metric on moduli space for generalized
multi-black hole solutions of the following action
\beq
I= {1\over16\pi G}\int d^D\!x\,\sqrt{-g}\left [R-
{\gamma\over 2}(\partial
\phi)^2-{1\over 4}e^{-a\gamma \phi}F^2 \right]
\label{mbone}
\eeq
where $\gamma=2/(D-2)$ and
$F=dA$ is an ordinary two-form field strength for the
vector potential $A$. For this theory, multi-centered solutions
describing extremal electrically
charged black holes have been found for arbitrary values of
the dilaton-vector coupling $a$ \cite{shir}. In these solutions,
the metric may be written in isotropic coordinates as
\beq
ds^2=-U^{-2}(\vec{x})dt^2+U^{2/(D-3)}(\vec{x})\,d\vec{x}^2
\label{metone}
\eeq
with $U(\vec{x})=H(\vec{x})^{(D-3)/(D-3+a^2)}$
and $H$ is a harmonic function
\beq
H(\vec{x})=1+\sum_{i=1}^n{\mu_i\over(D-3)|\vec{x}-\vec{x}_i|^{D-3}}\ .
\label{harone}
\eeq
The vector one-form potential is
\beq
A=\pm\sqrt{2\beta}
{dt\over H(\vec{x})}
\label{potone}
\eeq
and the dilaton is given by
\beq
e^{-\phi/a}=H(\vec{x})^{\beta}
\label{dilone}
\eeq
where $\beta=(D-2)/(D-3+a^2)$.

For the calculation of the Manton metric in the low velocity limit, the
positions $\vec{x}_i$ are made time dependent. Then to obtain a
solution
which satisfies the equations of motion to $O(v)$, one must solve for
the off-diagonal components of the metric (\eg $g_{t{x^i}}$)
and the spatial components of the gauge field. This solution is
substituted
into the field theory action (with care taken to regulate various
terms), and an effective lagrangian valid up to $O(v^2)$ is obtained.
Shiraishi
performed these calculations to produce the following effective
lagrangian describing the interactions of $N$
extremally charged black holes:
\beqa
L&=&-\sum_{i=1}^N m_i + \sum_{i=1}^N {1\over 2}
m_i {v}_i^2\label{effact}\\
&&+ \left(2-{1\over\beta}\right)
{2\pi G\over(A_{D-2})^2}
\int d^{D-1}\!x\, H(\vec{x})^{2(\beta-1)}
\sum_{i\ne j}^N
m_im_j{|\vec{v}_i-\vec{v}_j|^2\vec{r}_i\cdot\vec{r}_j
\over r_i^{D-1}\, r_j^{D-1}}\nonumber
\eeqa
where $\vec{r}_i=\vec{x}-\vec{x}_i$, and $\vec v_i$ and
$\vec x_i$ are,
respectively, the velocity and position of the $i$'th black hole.
Also $m_i$ is the mass of the $i$'th black hole given by
\[
m_i={A_{D-2}\over8\pi G}\beta\,\mu_i
\]
where $A_{D-2}=2\pi^{(D-1)/2}/\Gamma((D-1)/2)$ is the area of
a unit $(D-2)$-sphere.
The first two terms in (\ref{effact}) correspond to
the expected free particle Lagrangian to $O(v^2)$, while the remaining
term is the interaction Lagrangian. As expected the
interaction terms vanish as
the relative velocities go to zero. In general, this contribution
is highly nonlinear involving up to $N$-body interactions.
Collecting all of the $O(v^2)$ terms yields a metric on the
moduli space of these $N$ black hole configurations \cite{manton}.

The above interactions simplify for two specific values of the
scalar-Maxwell coupling.
For extreme $a^2=D-1$ black holes (\ie for which $\beta=1/2$),
the effective Lagrangian
reduces to the free terms only \cite{shir,shirscat}
\[
{L=-\sum_{i=1}^N m_i + \sum_{i=1}^N {1\over 2}
m_i v_i^2\ .}
\]
In other words,
the leading order velocity-dependent (\ie $O(v^2)$)
dynamical force between the black holes is zero, and the
low-velocity scattering is trivial (at least to this order).
Thus one infers
that the metric on the moduli space of these
extreme black holes is flat. Also
for $a^2=1$ (\ie for $\beta=1$), the Lagrangian
(\ref{effact}) simplifies in that it only
involves two-body interactions. Thus the effective
Lagrangian is easily determined to be simply \cite{shirscat}
\[
L=-\sum_{i=1}^N m_i + \sum_{i=1}^N {1\over 2}
m_i v_i^2 + {G\over A_{D-4}}\sum_{i\ne j}^N
m_im_j{|\vec{v}_i-\vec{v}_j|^2
\over |\vec{x}_i-\vec{x}_j|^{D-3}}
\]
for which the scattering calculations are greatly simplified
\cite{shirscat}.

Now suppose we begin with one of Shiraishi's multi-centered solutions
describing extremal electrically charged black holes,
and apply the first and second oxidation procedures
of section \ref{isox}, $p$ and $q$ times, respectively ---
in Appendix \ref{app} we perform this calculation explicitly for
the static solutions in eqs.~(\ref{metone}-\ref{dilone}).
The result will be a solution describing parallel $(p+q)$-branes
carrying an electrical charge of the $(p+1)$-form potential
for the oxidized theory in $D+p+q$ dimensions with the action
\beq
\bar{I}= {1\over16\pi \bar{G}}\int d^{D+p+q}\!x\,\sqrt{-\bar{g}}
\left[\bar{R}-{\bg\over 2}(\partial\bphi)^2
-{1\over 2(p+2)!}e^{-\ba\bg \bphi}\bar{F}^2 \right]
\label{newaction}
\eeq
where $\bg=2/(D+p+q-2)$ and $\ba^2=
[(D+p+q-2)/(D-2)]a^2-((D-2)pq+(D-3)^2p+q)/(D-2)$.
It should be clear that the dynamical interactions of these
$(p+q)$-branes
are identical to that of the original black holes, and is still
described by the effective lagrangian (\ref{effact}).
This observation follows from the fact that
the field-dependence of a given brane configuration is identical
to that of the corresponding dimensionally
reduced solution, and depends entirely on
the coordinates transverse to the $p$-brane.
The calculation which would verify this claim would begin
by oxidizing the non-static black hole
solutions which were explicitly calculated to leading order
in the velocity expansion by Shiraishi \cite{shirscat}.
Following the Manton method, these solutions are then to be
substituted into the field theory action, but we have seen in
sect.~\ref{isox} that the action
is unchanged by the oxidation procedure up to an
overall factor of the compactification volume.
Hence the effective Lagrangian density per unit $(p+q)$-volume would
be precisely the same as the Lagrangian (\ref{effact}) for the black
holes:
\beqa
\bar{L}&=&-\sum_{i=1}^N m_i + \sum_{i=1}^N {1\over 2}
m_i {v}_i^2\label{effactb}\\
&&+ \left(2-{1\over\beta}\right)
{2\pi \bar{G}\over(A_{D-2})^2}
\int d^{D-1}\!x\, H(\vec{x})^{2(\beta-1)}
\sum_{i\ne j}^N
m_im_j{|\vec{v}_i-\vec{v}_j|^2\vec{r}_i\cdot\vec{r}_j
\over r_i^{D-1}\, r_j^{D-1}}\ .\nonumber
\eeqa
Now $m_i$ is precisely the mass density of the $i$'th brane
with $m_i=A_{D-2}\beta\mu_i/(8\pi\bar{G})$, as per the discussion
in sect.~\ref{masse}.
In terms of $\ba$, one has $\beta=(D+p+q-2)/[(p+1)(D+q-3)+\ba^2]$.
As a result, the parallel $(p+q)$-branes interact in an identical
manner as their dimensionally reduced point-like ``descendants''.
Of course, by taking the dual form potential in Shiraishi's solutions,
one arrives at the same conclusion for
``magnetically'' charged $(p+q)$-branes.

Just as the black hole dynamics simplified in certain special cases,
so too will that of the $(p+q)$-branes. Only two-body interactions
arise for $\beta=1$ which now corresponds to $\ba^2=1-p(D+q-4)$. Here
the right-hand side is always negative --- recall that $D\ge4$ and $q\ge0$
--- unless $p=0$ in which case $\ba^2=1$. This situation then corresponds
to ``isotropic'' but not boost invariant $q$-branes carrying a
conventional $U(1)$ electric charge. The other case of interest is
$\beta=1/2$ for which the $O(v^2)$ interaction vanishes. This corresponds
to $\ba^2=4-(p-1)(D+q-5)$.
This latter condition can be satisfied in a wide number of cases.

As one particular example consider the cases with $q=0$ for which
$\ba^2=4-(p-1)(D-5)$. These solutions correspond to a particular class
of isotropic ``electric'' $p$-branes with $\kappa$-symmetry
\cite{report}.\footnote{Note that these solutions do not exhaust
the class of all $\kappa$-symmetric $p$-branes,
since the latter may in general involve more than one scalar and
antisymmetric tensor.} These (and in fact all of the oxidized) electrically
charged solutions are singular. Hence they
may be regarded as arising when the fields
of (\ref{finact}) are coupled to a source which in this particular
case is governed by a
$(p+1)$-dimensional supersymmetric sigma-model action \cite{sigma}
 whose kinetic and Wess-Zumino terms appear
with a relative coefficient fixed by the requirement of
$\kappa$-symmetry.
The role of the source action to the dynamics is to track the moving
sources with appropriate $\delta$-functions.
Then, in a standard notation \cite{report},
such solutions have $a^2=2(D'-2)-d'{\tilde d}'$, where $D'=D+p$ is the
total spacetime dimension, $d'=p+1$ is the dimension of the $p$-brane
world-volume, and ${\tilde d}'=D'-d'-2$ is the dimension of the
world-volume of the dual brane -- recall we have set $q=0$.
Following
the first oxidation/reduction procedure,
these $\kappa$-symmetric solutions reduce to $\kappa$-symmetric
solutions, and, ultimately, to black holes in $D=D'-p$ dimensions with
$a^2=D-1$, which, as we have seen above, were shown by Shiraishi to
scatter trivially. Using his results then, we have shown that
all of these $\kappa$-symmetric solutions
scatter trivially in the low-velocity limit. This has been
checked directly in a straightforward but tedious computation.
By duality considerations, trivial scattering holds for the solitonic,
``magnetic'' $p$-branes dual to the above, singular $p$-branes
\cite{report}.

\section{Discussion} \label{discuss}

Above we have derived a general effective Lagrangian
density (\ref{effactb}) describing the
slow motion scattering of arbitrary extremal $p$-branes. Using the
the double dimensional reduction/oxidation procedures of
sect.~\ref{isox}, we have shown that the $p$-brane dynamics
is identical to that of the corresponding dimensionally reduced
black holes. Because of our effective action is derived through
reduction/oxidation, it only describes the motion of parallel
$p$-branes moving in directions orthogonal to the surfaces of the
branes. For the conventional isotropic $p$-branes, this in fact
describes the entire moduli space of the interacting branes
because of the boost invariance of the solutions in the directions
parallel to the branes. For the more general anisotropic
$(p+q)$-branes considered in the Appendix, one might also consider
the motion of these branes in the $z^\ell$ directions --- \ie
boosting one of these solutions in a direction parallel to the $z^\ell$
produces a new brane which will not sit in equilibrium with one of the
original unboosted branes.

The most remarkable
result is that to leading order in the velocity expansion the
interactions vanish for all $\kappa$-symmetric $p$-branes within
the class of solutions that we are considering.
Such a flat metric on moduli space has previously
been found for $H$-monopoles \cite{monscat},  fundamental strings
\cite{stscat} and $D=10$ fivebranes \cite{fbscat}, all of which fall
into this
class of $\kappa$-symmetric solutions. More recently, Bachas
\cite{bachas}
has shown that for toroidal compactifications, $D$irichlet-branes (see
\cite{joe} and references therein) also have a flat metric.

The flat metric and consequent trivial dynamics for the
$\kappa$-symmetric solutions is a somewhat surprising result,
and is probably connected with the existence of flat directions
in the superpotentials associated with the underlying
$\kappa$-symmetric theories. Another possible answer is that
only these solutions preserve half the spacetime
supersymmetries in any supersymmetric embedding,
and inevitably this will also
constrain the dynamics considerably. For example,
if we embed the four-dimensional black holes in $N=8$,
$D=4$ supergravity, only the $\kappa$-symmetric
$a=\sqrt{3}$ black hole preserves four of the spacetime
supersymmetries.

The oxidation procedures of sect.~\ref{isox} can be applied to produce
a wide class of $p$-brane solutions given some family of black hole
solutions to an action of the form (\ref{finact})
or (\ref{newaction}) with arbitrary values
of the dilaton coupling constant $a$.
Shiraishi's solutions \cite{shir} involve {\it extremal} black
holes with electric charge, as required for the static zero-force condition
to hold. We use these as the basis for constructing a corresponding
family of isotropic (and anisotropic) brane solutions in Appendix
\ref{app}. Gibbons \cite{gibbons} has found general, \ie non-extremal,
static solutions describing a single black hole for the action
(\ref{finact}), again for arbitrary $a$.
One can just as easily apply the oxidation procedures of sect.~\ref{isox}
to these solutions to generate an even more
general class of isotropic (and anisotropic)
$(p+q)$-brane solutions of the action (\ref{newaction}). Typically
the nonextremal ``isotropic'' $p$-branes (setting $q=0$) will be isotropic
in the spatial brane directions, but they will not have the boost invariance
found in the extremal case. On the other hand, generically these
nonextremal branes produced by oxidization will have a real
event horizon rather than the singularities appearing in the
extremal case. While in many instances these oxidized branes can be thought
of as arising in a supersymmetric theory, they would usually not
preserve any of the supersymmetries. For this reason, these general
branes may ultimately be of less interest than the extremal solutions.
Note that in the extremal case, supersymmetry alone is not enough
to eliminate interactions once the branes are in motion. It is only
for the maximally supersymmetric, \ie the $\kappa$-symmetric, solutions
that the interactions vanish.

Note also that when the $p$-branes arise in
supersymmetric theories, the total amount of supersymmetry preserved
by a given $p$-brane is unaffected by the oxidation/reduction procedures
we have presented. This can be seen from the fact that these procedures
essentially do not change the field content of the original theory.
For example, the $a=\sqrt{3}$ black hole \cite{hmono}
preserves half the spacetime
supersymmetries in the $N=4, D=4$ theory, and continues to preserve
half the supersymmetries when oxidized to an anisotropic sixbrane in
the $N=1, D=10$ theory.

\section*{Acknowledgments}
This research was supported by NSERC of Canada and Fonds FCAR du
Qu\'ebec. R.R.K. was supported by a World Laboratory Fellowship.

\appendix

\section{Extremal $(p,q)$-branes} \label{app}

Let us begin with the action (\ref{mbone}) and Shiraishi's
black hole solutions described in eqs.~(\ref{metone}--\ref{dilone}).
Oxidizing this system $p$ times via the first procedure described in
section \ref{isox}, which then increases the form degree of the potential
in each step, the action becomes
\[
\hat{I}= {1\over16\pi \hat{G}}\int d^{D+p}\!x\,\sqrt{-\hat{g}}\left [\hat{R}-
{\hg\over 2}(\partial\hphi)^2
-{1\over 2(p+2)!}e^{-\ha\hg \hphi}\hat{F}^2 \right]
\]
where $\hg=2/(D+p-2)$ and $\ha^2=[(D+p-2)/(D-2)]\,a^2-(D-3)^2p/(D-2)$.
Shiraishi's black hole solutions have become $p$-brane solutions with
the addition of $p$ coordinates $y^a$ running parallel to the surface
of the brane. The vector potential becomes a $(p+1)$-form potential
which may be written as
\[
\hat{A}=\pm {\sqrt{2\beta}\over
H(\vec{x})}\ dt\,dy^1\ldots dy^p
\]
with the same harmonic function as in eq.~(\ref{harone})
\beq
H(\vec{x})=1+\sum_{i=1}^n{\mu_i\over(D-3)|\vec{x}-\vec{x}_i|^{D-3}}
\label{hartwo}
\eeq
and in terms of $\ha$, $\beta=(D+p-2)/[(p+1)(D-3)+\ha^2]$.
Similarly the dilaton is
\[
e^{-\hphi/\ha}=H(\vec{x})^\beta
\]
and the metric may be written
\[
d\hat{s}^2=\hat{U}^{-2}(\vec{x})\,(-dt^2+d\vec{y}^2)
+\hat{U}^{2{p+1\over D-3}}(\vec{x})\,d\vec{x}^2
\]
with $\hat{U}(\vec{x})=H(\vec{x})^{(D-3)/[(p+1)(D-3)+\ha^2]}$.
These solutions are in fact within the familiar class of static
``electrically'' charged $p$-brane solutions, which display
Lorentz invariance in the $(t,y^a)$ directions \cite{report}.

Now continue
oxidizing this system a further $q$ times using the second
procedure, so that this time the
the degree of the potential is held fixed.
The final action becomes
\beq
\bar{I}= {1\over16\pi \bar{G}}\int d^{D+p+q}\!x\,\sqrt{-\bar{g}}
\left[\bar{R}-{\bg\over 2}(\partial\bphi)^2
-{1\over 2(p+2)!}e^{-\ba\bg \bphi}\bar{F}^2 \right]
\label{mbthree}
\eeq
where $\bg=2/(D+p+q-2)$ and
\beqar
\ba^2&=&{D+p+q-2\over D+p-2}\,\ha^2-{(p+1)^2q\over D+p-2}\\
&=&{D+p+q-2\over D-2}\,a^2-{(D-2)pq+(D-3)^2p+q\over D-2}\ .
\eeqar
The final solutions are now $(p+q)$-brane solutions with a
further $q$ coordinates $z^\ell$ also running parallel to the surface
of the brane. The vector potential remains a $(p+1)$-form potential
which may be written as
\beq
\bar{A}=\pm {\sqrt{2\beta}\over
H(\vec{x})}\ dt\,dy^1\ldots dy^p\label{potthree}
\eeq
again with the same harmonic function (\ref{hartwo})
while in terms of $\bar{a}$, one has
$\beta=(D+p+q-2)/[(p+1)(D+q-3)+\ba^2]$. The dilaton becomes
\beq
e^{-\bphi/\ba}=H(\vec{x})^\beta
\label{dilthree}
\eeq
and the metric may be written as
\beq
d\bar{s}^2=\bar{U}^{-2}(\vec{x})\,(-dt^2+d\vec{y}^2)
+\bar{U}^{2{p+1\over D+q-3}}(\vec{x})\,(d\vec{x}^2+d\vec{z}^2)
\label{metthree}
\eeq
with $\bar{U}(\vec{x})=H(\vec{x})^{(D+q-3)/[(p+1)(D+q-3)+\ba^2]}$.
In this case the $(p+q)$-brane solutions are
Lorentz invariant in the $(t,y^a)$ directions. However, this
symmetry does not extend to include the coordinates $z^\ell$.
Note that there is no rotational symmetry between the $x^i$ and
$z^\ell$ directions because all of the fields and the geometry
depends on $x^i$ but not $z^\ell$.

Note that there is an alternative derivation for these particular solutions
as follows: If one attempts to construct a static $p$-brane solution
of the action (\ref{mbthree}), with the fields (\ref{potthree}-\ref{metthree})
as an ansatz, one would find that the
function $H$ must satisfy the Laplace's equation in a
$(D+q)$-dimensional flat space described by $\lbrace x^i,z^\ell
\rbrace$ (and also $H$ must
asymptotically approach one). If one chooses
this function to be independent of the $q$ coordinates $z^\ell$,
one is lead to the solution described above, and, in fact, since the
entire
solution is completely independent of $z^\ell$, one has actually constructed
a solution describing a $(p+q)$-brane. In fact, this approach also allows
for the construction of even more exotic solutions with non-parallel
branes.

By taking the dual form potential above,
one could also consider ``magnetically'' charged $(p+q)$-branes.
Completely new nonextremal anisotropic $(p+q)$-brane solutions
could be produced by applying the same construction described
above to the general black holes solutions of the action
(\ref{mbone}) presented in \cite{gibbons}. An example of this sort of
anisotropic $(p+q)$-brane is the $D=10$ uplifting of the (electrically
or magnetically charged)
$H$-monopole (or alternatively the uplifting of the extremally charged
Kaluza-Klein black hole)
\cite{hmono}, which has the structure of an anisotropic sixbrane with
$p=5$ and $q=1$.

As a final note, we
remark that one would refer to the above solutions as anisotropic
since the $y^a$ and $z^\ell$ directions are distinct.
However the solutions
with $p=0$ and $q\ne0$ are in fact isotropic in all of the
spatial $q$-brane solutions, even though they are not boost invariant.

\end{document}